\documentclass[twoside]{dis08}
\usepackage[latin1]{inputenc}
\usepackage[dvips]{graphicx,epsfig,color}
\usepackage{wrapfig,rotating}
\usepackage{amssymb,amsmath,array}

\newcommand{\ke}{K_{e2}}
\newcommand{\km}{K_{\mu 2}}

\begin{document}
\title{Searches for Physics beyond the Standard Model with Kaons
at NA48 and NA62 at CERN}

\author{Evgueni Goudzovski
%
%
\vspace{.3cm}\\
%
University of Birmingham - School of Physics and Astronomy \\
Edgbaston, Birmingham, B15 2TT - UK
}

\maketitle

\begin{abstract}
The ratio $R_K = \Gamma(K^{\pm}\to e^{\pm}\nu(\gamma)) /
\Gamma(K^{\pm}\to\mu^{\pm}\nu(\gamma))$ provides a powerful probe of
the structure of weak interactions. It is calculated with very high
precision within the Standard Model, but corrections due to the
presence of New Physics could be in a few percent range. Development
of NA48/NA62 method based on test data samples of 2003-04, and the
status of analysis based on a dedicated 2007 run are discussed. A
proposal to measure the ultra rare decay $K^+\to\pi^+\nu\bar\nu$ at
the CERN SPS is also presented.

\end{abstract}

\section*{Introduction}

The present CERN program in experimental kaon physics is represented
by the NA48 series of experiments at the SPS accelerator (physics
runs in 1997--2004), and its continuation NA62 (the first physics
run carried out in 2007).

Several NA48/NA62 activities aiming at the search for phenomena
beyond the Standard Model (SM) are discussed in the present paper.
These are 1) precise testing of lepton universality by measurement
of $R_K=\Gamma(K^\pm\to
e^\pm\nu(\gamma))/\Gamma(K^\pm\to\mu^\pm\nu(\gamma))$ based on test
data samples collected in 2003 and 2004 and a dedicated run of 2007;
2) designing an experiment to measure the branching ratio of a very
rare kaon decay $K^+\to\pi^+\nu\bar\nu$.

\boldmath
\section{Lepton universality test by measurement of $R_K$}
\unboldmath

\subsection{Motivation}

The $V-A$ structure and lepton universality are the two cornerstones
of the current description of weak interactions. The observation of
the $\pi^+\to e^+\nu$ decay at CERN in 1958 followed by later
measurements of $R_\pi=\Gamma(\pi^\pm\to e^\pm\nu(\gamma))/
\Gamma(\pi^\pm\to\mu^\pm\nu(\gamma))$, $R_K=\Gamma(K^\pm\to
e^\pm\nu(\gamma))/\Gamma(K^\pm\to\mu^\pm\nu(\gamma))$ and
$R_\tau=\Gamma(\tau^\pm\to e^\pm\nu_e
\nu_\tau)/\Gamma(\tau^\pm\to\mu^\pm\nu_\mu\nu_\tau)$ provided
verifications of the $V-A$ theory, confirming the suppression of
electronic decay modes due to helicity conservation.

The ratio $R_K$ can be predicted with a good accuracy in terms of
fundamental parameters due to a large degree of cancellation of the
hadronic uncertainties. By convention, the inner bremsstrahlung part
of the $K^\pm\to\ell^\pm\nu\gamma$ process is included into $R_K$,
while the structure dependent part (which is difficult to predict
theoretically) is not. Within the SM,
\begin{displaymath}
\label{Rdef} R_K=R_{\mathrm{tree}}(1 + \delta R_{\mathrm{QED}}) =
\left(\frac{m_e}{m_\mu}\right)^2
\left(\frac{m_K^2-m_e^2}{m_K^2-m_\mu^2}\right)^2 (1 + \delta
R_{\mathrm{QED}})= (2.477 \pm 0.001)\times 10^{-5},
\end{displaymath}
where $\delta R_{\mathrm{QED}}=-3.8\%$ is an electromagnetic
correction~\cite{cirigliano}. The factor $(m_e/m_\mu)^2$ accounts
for the helicity suppression of the $\ke$ mode.

A recent theoretical study~\cite{masiero} pointed out that, due to
the helicity suppression of $R_K$ in the SM, lepton-flavour
violating effects arising in super-symmetric extensions can induce
sizable violations of the $\mu-e$ universality, shifting $R_K$ from
the SM value by a relative amount that can be in the percent range,
without contradicting any other presently known experimental
constraints. The current world average $R_K=(2.45\pm 0.11)\times
10^{-5}$~\cite{pdg} is compatible with the SM prediction, being
however far from its current level of accuracy.

\subsection{Development of NA48/NA62 experimental method}

\begin{figure}[tb]
\vspace{-3mm}
\begin{center}
{\resizebox*{\textwidth}{!}{\includegraphics{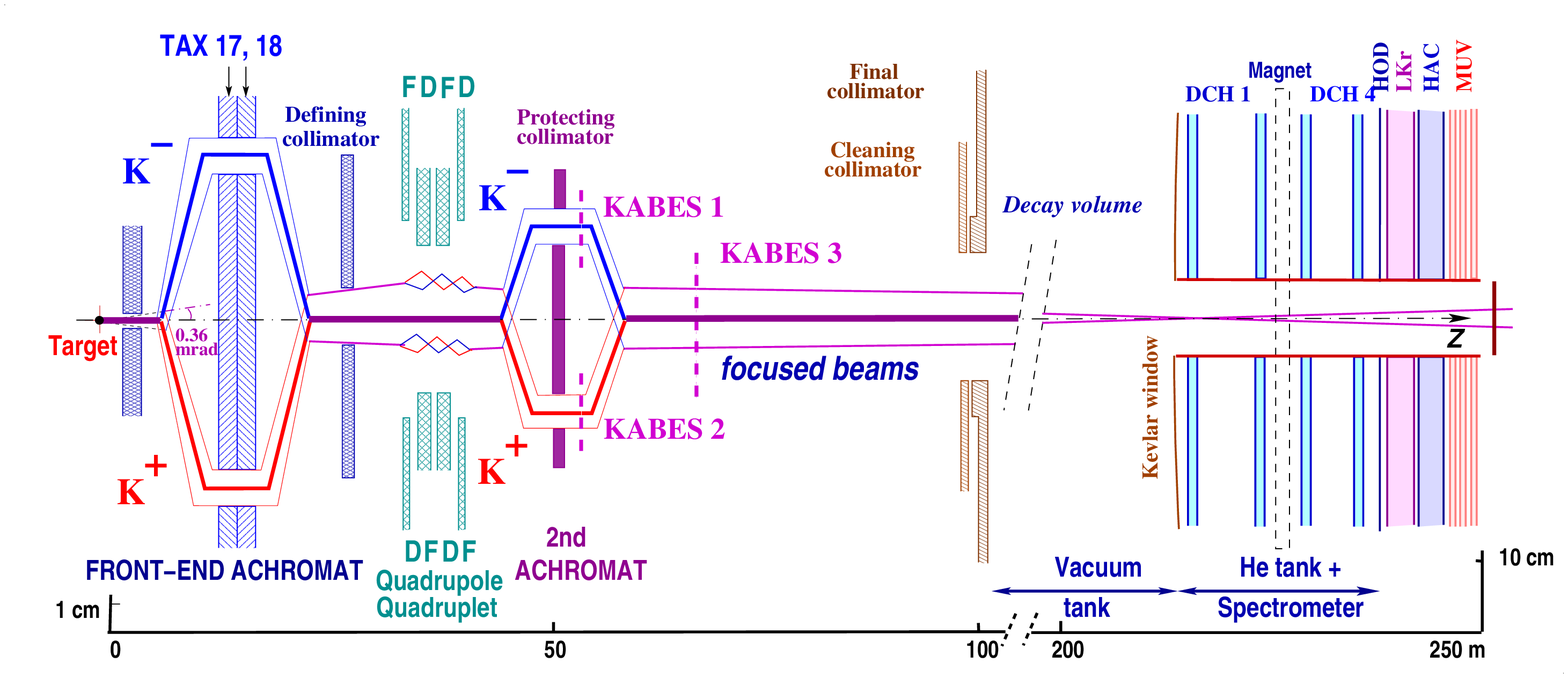}}}
\end{center}
\vspace{-6mm} \caption{Schematic lateral view of the NA48/2 beam
line (TAX17,18: motorized beam dump/collimators used to select the
momentum of the $K^+$ and $K^-$ beams; FDFD/DFDF: focusing set of
quadrupoles, KABES1--3: kaon beam spectrometer stations), decay
volume and detector (DCH1--4: drift chambers, HOD: hodoscope, LKr:
EM calorimeter, HAC: hadron calorimeter, MUV: muon veto). Vertical
scales differ in the two parts of the figure.} \label{fig:beams}
\end{figure}

The NA48 program dedicated to study of the charged kaon decays,
known as the NA48/2 experiment, is based on the physics runs of
2003--04. The experiment was designed to excel in charge asymmetry
measurements~\cite{ba07}, and is based on simultaneous $K^+$ and
$K^-$ beams produced by 400 GeV/$c$ primary SPS protons interacting
with a beryllium target. The layout of the beams and detectors is
shown schematically in Fig.~\ref{fig:beams}. Charged particles with
momentum $(60\pm3)$ GeV/$c$ are selected by an achromatic system of
four dipole magnets with zero total deflection (`achromat'), which
splits the two beams in the vertical plane and then recombines them
on a common axis.

The beams then enter a fiducial decay volume housed in a 114 m long
cylindrical vacuum tank. The decay volume is followed by a magnetic
spectrometer consisting of four drift chambers, a trigger
scintillator hodoscope, a liquid krypton electromagnetic
calorimeter, a hadron calorimeter, and a muon detector. Further
details on the experimental setup can be found
elsewhere~\cite{fa07}.

During a part of the 2003 run (4 weeks of data taking), a two level
$\ke$ trigger chain suitable for $R_K$ measurement was implemented
along with the main triggers. The first level (L1) consisted of a
time coincidence of hits in the two planes of the HOD (so called
$Q_1$ signal) with an energy deposition of at least 10 GeV in the
calorimeter. The second level (L2) consisted of a requirement that
the missing mass $m_X$ computed using the track momentum measured by
the spectrometer in the hypothesis of $K\to\pi X$ decay is below the
$\pi^0$ mass. The $\ke$ trigger was downscaled by a factor ranging
from 20 to 40 during the data taking. The $\km$ trigger consisted of
the $Q_1$ condition only downscaled by a factor of $10^4$.

The 2003 data sample of 4670 $\ke^\pm$ candidates with 12\%
estimated background was used to establish the analysis strategy.
However the data quality appeared to be affected by high
inefficiency of the L2 trigger, measured to be about 15\% (while the
L1 inefficiency was measured to be below 1\%). As a consequence, the
uncertainty of the 2003 result received a $\delta R_K/R_K=0.8\%$
contribution from the L2 trigger inefficiency.

To overcome the above limitation, a dedicated 56 hour run was taken
in 2004 with a simplified $\ke$ trigger consisting of the L1 part
only without downscaling, the main trigger chains disabled, and beam
intensity reduced by a factor of 4 with respect to the nominal
value. Analysis of those data led to further refinements of the
experimental approach.

Event selection is based on (1) kinematic criteria: $K_{\ell 2}$
candidates are required to have squared missing mass $M_{\rm
miss}^2(\ell)=(P_K-P_\ell)^2$ compatible to zero, where $P_K,P_\ell$
($\ell = e,\mu$) are 4-momenta of kaon (beam average assumed) and
lepton (electron or muon mass assumed); (2) particle identification
based on the ratio $E/p$ of track energy deposition in the
calorimeter to track momentum: particles with $E/p>0.95$ ($E/p<0.2$)
are identified as electrons (muons). The use of the identification
criteria necessitates a detailed study of the calorimeter response:
particle identification probabilities are measured with the data.

Analysis of the 2004 data sample containing 3930 $\ke$ candidates
with 14\% estimated background was focused on the technique of
background subtraction in the $\ke$ sample. The major contribution
(12\%) is due to $\km$ decays with muons releasing almost all their
energy in the calorimeter by `catastrophic' bremsstrahlung. It
contributes at high lepton momenta only, in the region of poor
$\ke$/$\km$ kinematical separation. Its subtraction requires a
measurement of the probability $P(\mu\to e)$ of the (rare) muon
misidentification as electron due to having $E/p>0.95$, which
requires selection of a very clean muon sample free of electron
admixture. In 2003--04 data taking conditions, only the kinematical
region of low lepton momentum $p<35$~GeV/$c$, i.e. good $\ke$/$\km$
kinematical separation, in which kinematical selection of a pure
$\km$ sample is possible, is accessible for such measurement. It was
measured in this region: $P(\mu\to e)=4\times10^{-6}$. However
theory suggests growth of `catastrophic' bremsstrahlung cross
section as a function of muon momentum~\cite{ke97}, which impedes
reliable $\km$ background subtraction. As a consequence, the
precision of the 2004 result is limited by a corresponding
uncertainty of $\delta R_K/R_K=1.6\%$.

\boldmath
\subsection{A dedicated $K^\pm\to e^\pm\nu$ run in 2007}
\unboldmath

A dedicated $\ke$ run of the NA62 experiment was carried out in
June--October of 2007 with the aim of reaching an improved level
accuracy of better than $\delta R_K/R_K=0.5\%$. The data taking
conditions were optimized with respect to 2004 using the past
experience as follows.
\begin{itemize}
\item $M_{\rm miss}^2$ resolution was improved by using narrow
band $K^\pm$ beams (2\% RMS vs 3\% in 2003--04), and larger
spectrometer magnet momentum kick (263 MeV/$c$ vs 120 MeV/$c$ in
2003--04, reaching lepton momentum resolution of $\delta p/p=0.47\%
\oplus 0.020\%p$, $p$ in GeV/$c$). Expected $\km$ contamination in
$\ke$ sample is 7\% vs 12\% in 2003--04.
\item Selection of pure muon samples in the whole analysis muon
momentum range was made possible by installing a $\sim 10X_0$ thick
lead wall (absorbing a large fraction of electron, but not muon,
energy deposition) between the two planes of the scintillator
hodoscope covering $\sim 20\%$ of the geometric acceptance.
Preliminary measurements of $P(\mu\to e)$ confirm the predicted
momentum dependence of this quantity.
\item A number of runs with
special conditions were carried out to address the measurements of
lepton misidentification probabilities, and the background induced
by the beam halo.
\end{itemize}
About $1.1\times 10^5$ $\ke$ candidates with $<10\%$ background were
collected, data analysis started.

\boldmath
\section{A proposal to measure $\rm{BR}(K^+\to\pi^+\nu\bar\nu)$ at CERN}
\unboldmath

The $K^+\to\pi^+\nu\bar\nu$ decay is a FCNC process with the SM
branching ratio computed to be $(7.83\pm0.82)\times
10^{-11}$~\cite{bu06}. Its unique theoretical cleanness owes to the
possibility of parameterizing the hadronic matrix element via the
experimentally well known ${\rm BR}(K^+\to\pi^0e^+\nu)$; theoretical
uncertainty is largely due to those of the CKM matrix elements.

The NA62 proposal to collect $\sim100$ decays with 10\% background
in 2 years of operation, as required to match the theory precision,
is based on the existing NA48 kaon beam line and infrastructure. The
R\&D started in 2006, and data taking is foreseen to start in 2011.
Decay in flight technique is chosen, reaching an acceptance of 10\%;
the beam line is required to provide $10^{13}$ kaon decays from
unseparated beam with 6\% kaon fraction in about 200 days of data
taking.

The experimental signature of the $K^+\to\pi^+\nu\bar\nu$ decay is a
single reconstructed track in the detector downstream the decay
volume in time coincidence with a kaon measured by the upstream beam
tracker. Kinematic selection is performed using $K$ momentum
measurement by a silicon pixel detector in the beam operating at
rate of 800 MHz, and $\pi$ momentum measurement by a magnetic
spectrometer consisting of straw chambers operating in vacuum.

In addition to kinematic rejection (which provides large rejection
factors for $K_{\mu2}$, $K_{2\pi}$ and $K_{3\pi}$ decays, but is not
sufficient for background suppression), the following techniques are
used to provide redundancy. Particle identification is based on
CEDAR differential Cherenkov counter~\cite{cedar} for kaons, and
RICH detector for pions. A series of photon detectors are used to
provide a hermetic photon veto with an inefficiency of
$\sim10^{-5}$: these are ring shaped scintillator fibre calorimeters
at large angles of 10--50 mrad, the existing NA48 liquid crypton
calorimeter at 1--10 mrad, and a set of small angle vetoes based on
shashlyk technology at low angles. Prototypes of the new detectors
were successfully tested at CERN and Frascati in 2006--2007; a
series of further tests is foreseen in 2008.


\begin{footnotesize}

\end{footnotesize}



\begin{thebibliography}{99}
\bibitem{url} Slides: \\
\verb$http://indico.cern.ch/contributionDisplay.py?contribId=67&sessionId=15&confId=24657$
\bibitem{cirigliano} V.~Cirigliano and I.~Rosell, JHEP {\bf 0710} 005 (2007).
\bibitem{masiero} A.~Masiero, P.~Paradisi and R.~Petronzio, Phys. Rev. {\bf D74} 011701 (2006).
\bibitem{pdg} W.-M.~Yao {\it et~al.} (PDG), J. Phys. {\bf G33} 1 (2006).
\bibitem{ba07} J.R.~Batley {\it et~al.}, Eur. Phys. J. {\bf C52} 875 (2007).
\bibitem{fa07} V.~Fanti {\it et~al.}, Nucl. Inst. Methods {\bf 574} 433 (2007).
\bibitem{ke97} S.R.~Kelner, R.P.~Kokoulin and A.A.~Petrukhin, Phys.
Atom. Nucl. {\bf 60} 576 (1997).
\bibitem{bu06} F.~Mescia and C.~Smith, Phys. Rev. {\bf D76} 034017 (2007) and
references [1,2] therein.\newline See also
\verb$http://www.lnf.infn.it/wg/vus$
\bibitem{cedar} G.~Bovet {\it et~al.}, CERN report CERN 82-12 (1993).
\end{thebibliography}
\end{document}